# Superfluorescence from Lead Halide Perovskite Quantum Dot Superlattices

Gabriele Rainò[1,2,3*], Michael A. Becker[3,4*], Maryna I. Bodnarchuk[2], Rainer F. Mahrt[3], Maksym V. Kovalenko[1,2], and Thilo Stöferle[3]

[1]Institute of Inorganic Chemistry, Department of Chemistry and Applied Bioscience, ETH Zurich, 8093 Zurich, Switzerland.

[2]Laboratory of Thin Films and Photovoltaics, Empa – Swiss Federal Laboratories for Materials Science and Technology, 8600 Dübendorf, Switzerland.

[3]IBM Research – Zurich, Säumerstrasse 4, 8803 Rüschlikon, Switzerland.

[4]Optical Materials Engineering Laboratory, ETH Zurich, 8092 Zurich, Switzerland.

[*]These authors contributed equally to this work.

**An ensemble of emitters can behave significantly different from its individual constituents when interacting coherently via a common light field. After excitation, collective coupling gives rise to an intriguing many-body quantum phenomenon, resulting in short, intense bursts of light: so-called superfluorescence[1]. Because it requires a fine balance of interaction between the emitters and their decoupling from the environment, together with close identity of the individual emitters, superfluorescence has thus far been observed only in a limited number of systems, such as atomic and molecular gases[2] and semiconductor crystals[3–7], and could not be harnessed for applications. For colloidal nanocrystals, however, which are of increasing relevance in a number of opto-electronic applications[8], the generation of superfluorescent light was precluded by inhomogeneous emission broadening, low oscillator strength, and fast exciton dephasing. Using caesium lead halide (CsPbX$_3$, X = Cl, Br) perovskite nanocrystals[9–12] that are self-organized into highly ordered three-dimensional superlattices[13,14] allows us to observe key signatures of superfluorescence: red-shifted emission with more than ten-fold accelerated radiative decay, extension of the first-order coherence time by more than a factor of four,**

**photon bunching, and delayed emission pulses with Burnham–Chiao ringing behaviour[15] at high excitation density. These mesoscopically extended coherent states can be employed to boost opto-electronic device performances[16,17] and enable entangled multi-photon quantum light sources[18-20].**

Spontaneous emission (SE) of photons, such as fluorescence commonly used in displays or lighting, occurs due to coupling excited two-level systems (TLS) to the vacuum modes of the electromagnetic field, effectively stimulated by its zero-point fluctuations[21]. In 1954, Dicke predicted[22] that an ensemble of $N$ identical TLS confined in a volume smaller than $\sim\lambda^3$ ($\lambda$ is the corresponding emission wavelength of the TLS) can exhibit coherent and cooperative spontaneous emission. This so-called superradiant emission results from the coherent coupling between individual TLS through the common vacuum modes, effectively leading to a single giant emitting dipole from all participating TLS. In the case when the excited TLS are initially fully uncorrelated, and the coherence is only established from spontaneously produced correlations due to quantum fluctuations rather than by coherent excitation, a superfluorescent (SF) pulse is emitted[1] (Figure 1). Both superradiant emission and coherent SF bursts are characterized by an accelerated radiative decay time $\tau_{SF} \sim \tau_{SE}/N$, where the exponential decay time $\tau_{SE}$ of the uncoupled TLS is shortened by the number of coupled emitters $N$. In addition, SF exhibits the following fundamental signatures (i) a delay or build-up time $\tau_D \sim \ln(N)/N$ during which the emitters couple and phase-synchronize to each other, and which corresponds to the time delay between the excitation and onset of the cooperative emission (Figure 1) and (ii) coherent Rabi-type oscillations in the time domain due to the strong light–matter interaction, known as Burnham–Chiao ringing[15,23].

SF was first observed in a dense gas of hydrogen fluoride[2], followed by a limited number of solid state systems, such as CuCl nanocrystals (NCs) embedded in a NaCl matrix[4], KCl crystals doped with peroxide anions ($O_2^-$) (ref. [3]) and some select semiconductor crystals[5,6]. However, the restrictions on the materials imposed by the requirement for high oscillator strength, small inhomogeneous line-broadening, together with little dephasing, make it difficult to exploit SF for solid-state applications, such as ultrafast light-emitting diodes or quantum sources of entangled photons. Colloidal semiconductor NCs or quantum dots (QDs) could fill this gap as they are inexpensive, easily processable, and a versatile material class already employed for advanced photonic applications[8,24]; however, up to now, they have not been able to match the stringent properties necessary for SF.

Here, we use colloidal NCs of caesium lead halide perovskites ($CsPbX_3$, X = Cl, Br, or I) that can be produced with narrow size dispersion and are known to exhibit moderate quantum confinement effects, resulting in narrow-band emission combined with exceptionally large oscillator strength from a bright triplet state[9,10,25]. In order to foster cooperative behaviour, we employ structurally well-defined, long-range ordered, and densely packed arrays of such NCs, known as superlattices, constructed by means of solvent-drying-induced spontaneous assembly[13,14,26,27]. Similarly, regular arrays of II-VI semiconductor NCs have been used to obtain collective effects in the electronic domain, *i.e.,* band-like transport[17]. Figure 2a outlines the superlattice formation (see also Supplementary Information(SI)), using a solution of highly monodispersed $CsPbBr_3$ NCs with a mean size of 10 nm and standard size-deviation of less than 10% (Figure 2b). In the self-assembly process, cubic individual superlattice domains are formed (*i.e.*, supercrystals), each consisting of up to several millions of NCs. Optical microscopy (Figure 2c) reveals superlattices with a lateral size of up to 5 µm, randomly distributed in a uniform film on a 5 × 7 mm sample (Figure 2d). Transmission electron microscopy confirms that highly ordered superlattices consist of well-separated individual NCs (Figure 2e). More details of the self-assembly process are reported in the SI.

Figure 3a displays the photoluminescence (PL) spectrum of a single $CsPbBr_3$ superlattice at 5 K exhibiting two peaks. The high-energy emission peak coincides with the single-peaked spectrum of an ensemble of $CsPbBr_3$ NCs in an amorphous (glassy) film and is therefore assigned to non-coupled QDs. In addition, a narrow, red-shifted emission peak appears in superlattices, which we assign to the cooperative emission of QDs. We can exclude that this feature, which is typically red-shifted by 70−90 meV from the uncoupled QD emission, originates from the emission from trions, bi-excitons, or multi-excitons, because their energy shifts are reportedly 10−20 meV (refs. [10,11]). The number and interaction strength of coherently coupled QDs determine the magnitude of the energetic shift. In most superlattices, we observe a sub-structure in this red-shifted emission band, which we attribute to the presence of several, slightly different independent SF domains within the same individual superlattice.

A central feature of the cooperative emission is the modification of the radiative lifetime[22], as demonstrated experimentally with several quantum emitters[6,16]. In time-resolved PL decay measurements, we observe an accelerated PL decay of the SF emission peak in comparison to the PL decay of uncoupled QDs with $1/e$ decay times of $\tau_{SF} = 148 \text{ ps}$ and $\tau_{QD} = 400 \text{ ps}$, respectively, for an excitation density of 500 nJ/cm² per pulse (Figure 3b). In contrast to the predominantly mono-exponential decay of the uncoupled QDs, the SF emission decay is approximated well by a stretched exponential[28], because the number of excited coupled emitters, and therefore the speed-up, varies during the decay. Furthermore, in contrast to the uncoupled QDs, the SF decay time is strongly dependent on excitation power (inset Figure 3b) because it scales with the coupling strength among the QDs, given by the intensity in the common light-field that effectively corresponds to a change in the number of coherently coupled QDs. When the spectrally and temporally integrated emission is fitted with a power law, we obtain an exponent of 1 (Extended Data Figure 1b), indicating that excitation density-dependent non-radiative decay channels (e.g. Auger recombination) are absent.

The cooperative emission process strongly influences the coherence of the emitted light. First-order correlation measurements of each of the two emission peaks by means of a Michelson interferometer allow us to monitor the interference pattern and therefore the phase coherence time (Figure 4a). The emission band of the uncoupled QDs exhibits 38 fs coherence time, best fitted with a Gaussian decay (Figure 4a, upper graph), typical of incoherent (thermal) light sources[29]. The emission from the coherently coupled QDs (Figure 4a, lower graph) displays a much longer coherence time with an exponential decay of 140 fs. For some superlattices, a Gaussian decay is observed (Extended Data Figure 3b), which might be attributed to number fluctuations within the coherent SF state[30].

Second-order coherence of the emitted light is evinced by the statistics of the photon arrival time on a detector[31]. Typical coherent light, as from a laser, shows a random distribution (Poissonian) of photon arrival times, while a single TLS exhibits photon antibunching (sub-Poissonian distribution). In contrast, the cooperative emission from coupled QDs leads to coherent multi-photon emission bursts. Figure 4 reports the second-order correlation function, $g^{(2)}(\tau) = \frac{\langle I(t)I(t+\tau)\rangle}{\langle I(t)\rangle^2}$ for both PL emission bands, where $I(t)$ is the signal intensity at time $t$. For the uncoupled QD emission (Figure 4b, upper graph), a flat $g^{(2)}(\tau) = 1$ is observed because the experimental temporal resolution (40 ps) is insufficient to resolve the expected thermal bunching[32]. The SF emission band, however, shows pronounced photon bunching (Figure 4b, lower graph) because the coherent coupling leads to the correlated emission of multiple photons within a short time interval. Photon bunching is only observable in superlattices with single or a few SF domains, *i.e.*, where no sub-structure is visible in the red-shifted emission band, because spectrally overlapping uncorrelated aggregated domains within the same superlattice reduce the bunching peak's visibility, as predicted by theory[33]. Yet, it is a robust effect that is observed with pulsed excitation and for mixed-halide (CsPbBr$_2$Cl, emitting at higher energies) QD superlattices too (see Extended Data Figure 3a and 4b, respectively). Remarkably, some superlattices with supposedly well-isolated coherently coupled QDs exhibit $g^{(2)}(\tau) > 2$ (inset Figure 4), similar to superthermal emission[31]. The exponential decay time of the second-order correlation is of the order of the radiative decay of the SF emission for low excitation densities ($\tau_{g^{(2)}} = 224 \text{ ps}$).

Very distinct characteristics of SF emission concern the time evolution of the emitted light under strong driving conditions. Figure 5a shows a streak camera image acquired at an excitation density of 1600 µJ/cm$^2$, where in addition to a drastically shortened radiative decay, a finite rise time and subsequent oscillations of the emission are observed. Quantitative analysis on spectrally integrated PL decay traces for various excitation power densities are shown in Figure 5b (for details see SI). As a function of the excitation density, the decay time shortens to below 8 ps (Figure 5c, upper panel). The peak intensity increases super-linearly over three orders of magnitude (Figure 5c, middle panel), according to a power-law dependence with an exponent of $\alpha = 1.3 \pm 0.1$, deviating from the theoretically expected value of $\alpha = 2$ (ref. [7]) probably due to saturation effects[5]. Nevertheless, no significant quenching effects of the emission for high excitation powers were found, verifying that decay remains essentially radiative (Extended Data Figure 1d). Furthermore, a shortening of the SF delay time ($\tau_D$), after which the photon burst is emitted, is observed (Figure 5c, bottom panel). This characteristic of SF is a consequence of the time it takes for phase-locking the individual dipoles and scales with the number $N$ of excited coupled QDs according to $\tau_D \sim \frac{\log(N)}{N}$ (see SI).

As SF crucially depends on low decoherence and low inhomogeneous spread, it should be noted that SF coupling is strongly affected by the environment around the QDs (number of free ligands), the superlattice assembly, and by the quality of the QDs themselves. Thus, while a large fraction of the superlattices displays a red-shifted peak from the cooperative emission, the amount of photon-bunching and Burnham–Chiao ringing varied from superlattice to superlattice. However, experiments employing different batches of NCs and superlattice assemblies of $CsPbBr_3$ and $CsPbBr_2Cl$ NCs were consistently reproducible, but further optimization of the synthesis and assembly is likely to improve the yield of SF domains.

Our measurements reveal that coherent SF coupling can be achieved in long-range ordered self-assembled superlattices of fully inorganic $CsPbX_3$ perovskite NCs, resulting in strong emission bursts. Colloidal NCs and their assemblies have proven to be excellent building blocks for a large variety of opto-electronic devices, and these cooperative effects now allow modification of the opto-electronic properties beyond what is possible on the individual QD level with chemical engineering approaches. This opens up new opportunities for high-brightness and multi-photon quantum light sources, and could enable the exploitation of cooperative effects for long-range quantum transport and ultra-narrow tuneable lasers.

**Data availability.** The data that support the findings of this study are available from the corresponding authors upon reasonable request.

**Supplementary Information** is available in the online version of the paper.

**Acknowledgements**

We thank D. J. Norris, C. Schwemmer, D. Urbonas, and F. Scafirimuto for helpful discussions. M.A.B., M.V.K., T.S., R.F.M., and G.R. acknowledge the European Union's Horizon-2020 programme through the Marie-Sklodowska Curie ITN network PHONSI (H2020-MSCA-ITN-642656) and the Swiss State Secretariat for Education Research and Innovation (SERI). M.I.B. acknowledges financial support from the Swiss National Science Foundation (SNF Ambizione grant, Grant No. PZENP2_154287). M.V.K. acknowledges financial support from the European Research Council under the European Union's Seventh Framework Program (FP/2007-2013) / ERC Grant Agreement No. 306733 (NANOSOLID Starting Grant).

**Author contributions**

The work originated from ongoing interactions between G.R., M.V.K, R.F.M., and T.S. M.I.B. prepared the samples and performed their structural characterization. G.R., M.A.B., and T.S. performed the optical experiments, interpreted the data with input from R.F.M. G.R. and M.A.B. wrote the manuscript with input from all the co-authors. R.F.M., M.V.K., and T.S. supervised the work.

**Author information**

The authors declare no competing financial interests. Readers are welcome to comment on the online version of the paper. Correspondence and requests for materials should be

# Figures

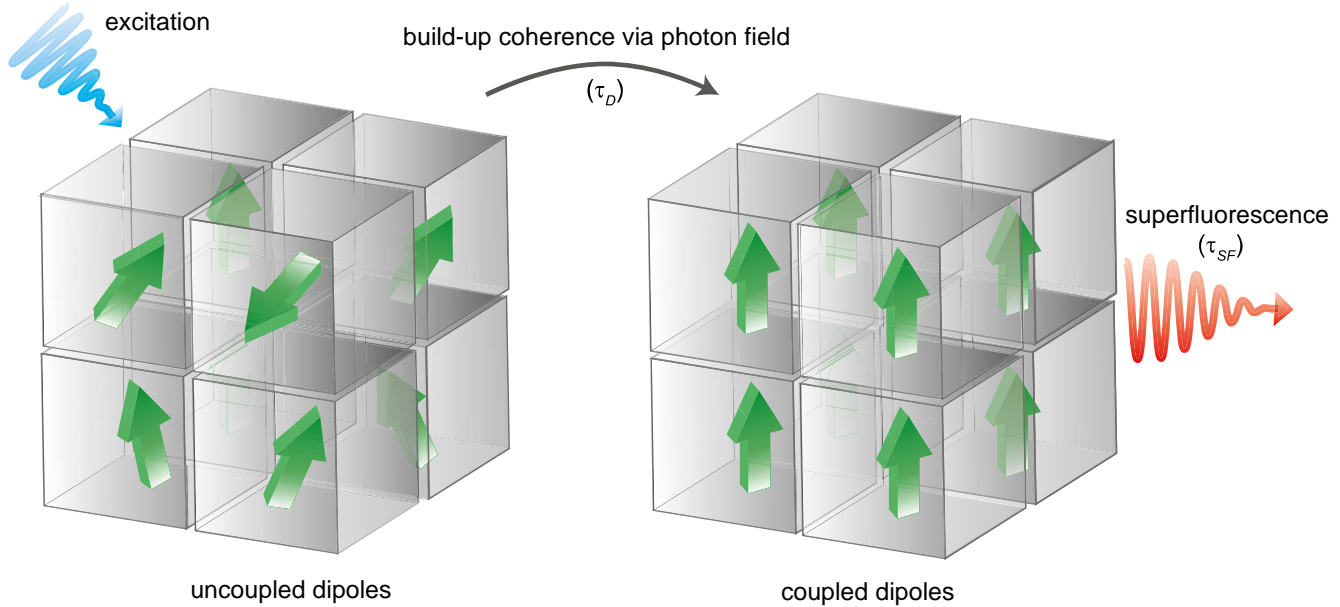

**Figure 1 | Schematic of the build-up process of SF.** An initially uncoupled ensemble of TLS (randomly oriented green arrows) is excited by a light pulse (blue arrow). After time $\tau_D$ their phases are synchronized (aligned green arrows) such that they cooperatively emit a SF light pulse (red arrow). Grey cubes represent long-range ordered self-assembled superlattices.

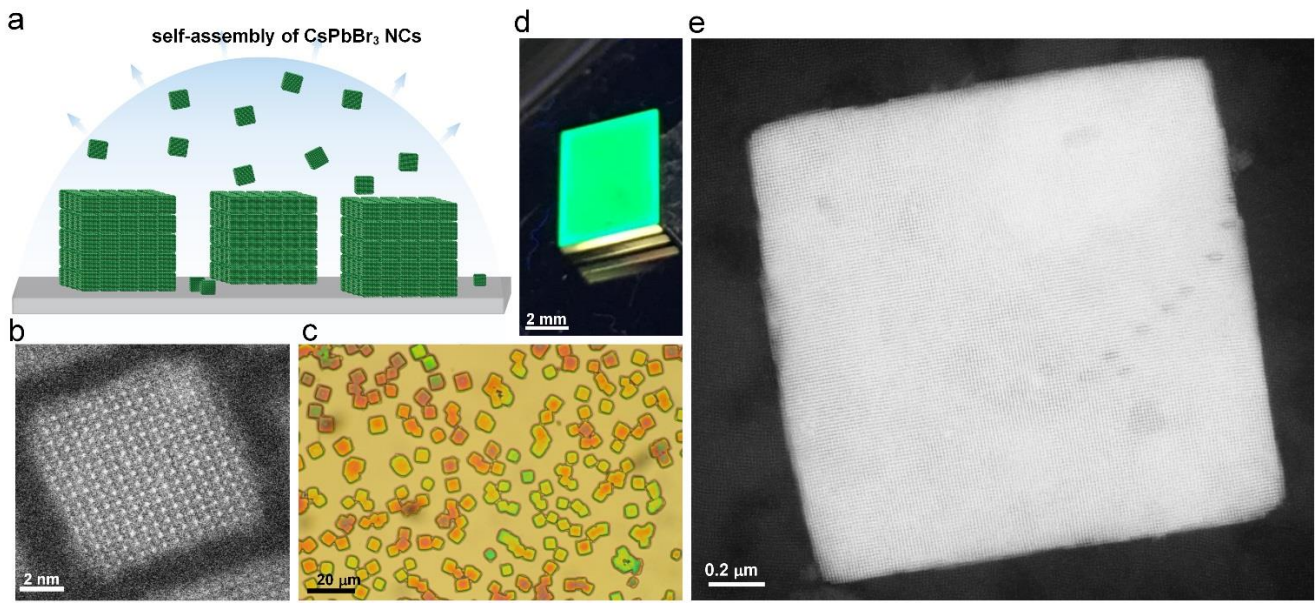

**Figure 2 | Formation of CsPbX$_3$ (X=Cl, Br) NC superlattices by drying-mediated self-assembly. a**, Illustration of the assembly process. **b**, High-resolution scanning transmission electron microscopy image (HAADF–STEM) of a single CsPbBr$_3$ NC. **c**, Optical microscope image and **d**, photograph (under UV light) of a layer of micron-sized, three-dimensional, cubic-shaped NC superlattices. **e**, HAADF–STEM image of a single superlattice of CsPbBr$_3$ NCs. The cubic shape of the individual perovskite NC building blocks is translated into the symmetry of the superlattice (simple cubic packing).

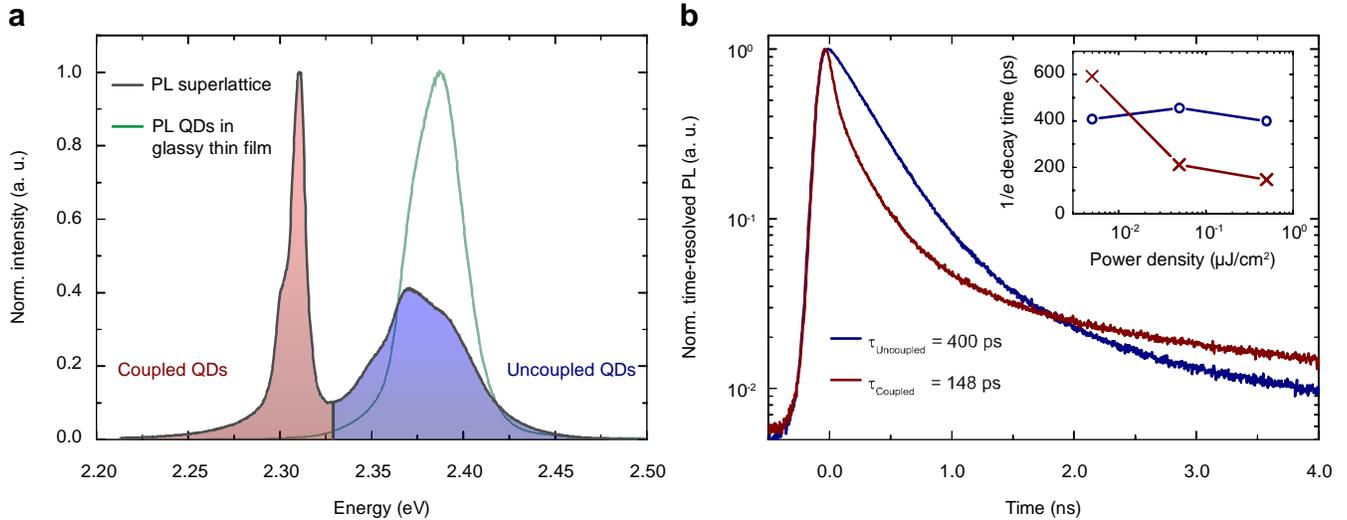

**Figure 3 | Optical properties of CsPbBr$_3$ NC superlattices. a**, PL spectrum of a single CsPbBr$_3$ superlattice (black solid line). The high-energy band is assigned to the emission of uncoupled QDs. The low-energy band is the result of the cooperative emission of coherently coupled QDs and is not present in glassy films of NCs (green solid line). **b**, Time-resolved PL decay of the two emission bands at 500 nJ/cm$^2$ excitation density after applying suitable spectral filters to separate the two components. The decay from the coherently coupled QDs is significantly faster than from the uncoupled ones. The inset shows the power-dependence of the 1/$e$-decay times of both components.

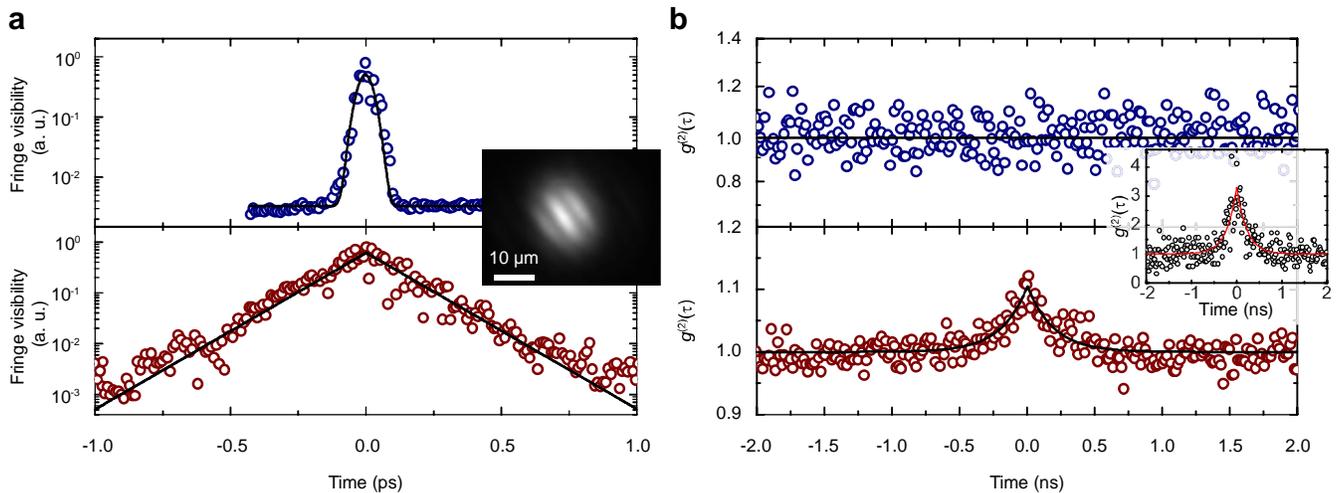

**Figure 4 | First- and second-order coherence properties. a**, First-order correlation of the two emission bands as obtained from the interference fringe visibility using a Michelson interferometer. The high-energy band of the uncoupled QDs has a very short phase coherence time (<40 fs, upper graph), whereas the red-shifted band from the coupled QDs is characterized by much longer phase coherence (140 fs, lower graph). The solid lines are fits to the data (see text). The inset shows an example of the real space interferogram. **b**, Second-order correlation function, $g^{(2)}(\tau)$, obtained with a Hanbury–Brown and Twiss setup in start–stop configuration. For the high-energy band (upper graph), a flat profile with $g^{(2)}(\tau) = 1$ is observed. The red-shifted emission band (lower graph) from the SF emission displays a pronounced bunching peak, characteristic of the correlated emission during a photon burst. The data are fitted to the function $g^{(2)}(\tau) = 1 + A \cdot exp(-(|\tau - \tau_0|/\tau_c)$ (solid lines). The inset shows an example of superbunching with $g^{(2)}(0) > 2$ that has been observed with some superlattices.

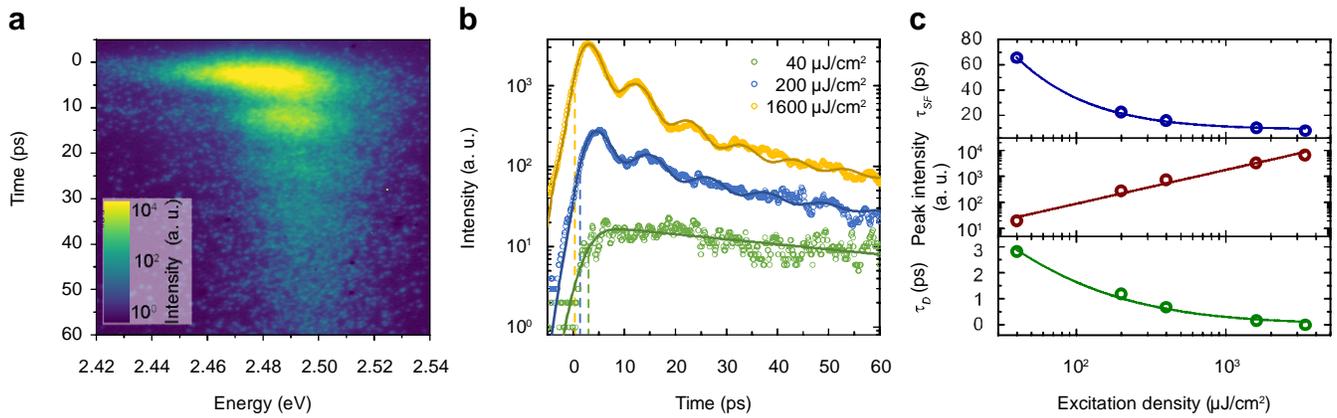

**Figure 5 | Burnham–Chiao ringing behaviour. a**, Streak camera image of SF dynamics obtained with high excitation density of 1600 µJ/cm². **b**, Extracted time-resolved emission intensity traces for three different excitation powers. Solid lines are best-fits to a model that employs a bi-exponential decay function with damped oscillations. **c**, *Top:* Effective SF decay (blue circles) as a function of the excitation power density fitted according to the SF model (solid blue line). *Middle:* Red circles represent the peak SF emission intensity that increases super-linearly with excitation power, corresponding to a power-law dependence with an exponent $\alpha = 1.3 \pm 0.1$ (solid dark-red line). *Bottom:* The extracted delay time $\tau_D$ (green circles) decreases at high excitation power due to the increased interaction among the emitters. The green solid line is the best fit according the model described in the SI.

# Supplementary Information

# Superfluorescence from Lead Halide Perovskite Quantum Dot Superlattices

Gabriele Rainò[1,2,3*], Michael A. Becker[3,4*], Maryna I. Bodnarchuk[2], Rainer F. Mahrt[3], Maksym V. Kovalenko[1,2], and Thilo Stöferle[3]

[1]Institute of Inorganic Chemistry, Department of Chemistry and Applied Bioscience, ETH Zurich, 8093 Zurich, Switzerland.

[2]Laboratory of Thin Films and Photovoltaics, Empa – Swiss Federal Laboratories for Materials Science and Technology, 8600 Dübendorf, Switzerland.

[3]IBM Research – Zurich, Säumerstrasse 4, 8803 Rüschlikon, Switzerland.

[4]Optical Materials Engineering Laboratory, ETH Zurich, 8092 Zurich, Switzerland.

[*]These authors contributed equally to this work.

**I) MATERIALS AND METHODS**

A) NANOCRYSTAL SYNTHESIS AND SUPERLATTICE FORMATION

**Synthesis of CsPbBr$_3$ nanocrystals**. In a 25 ml three-necked flask, PbBr$_2$ (69 mg, 0.188 mmol, Aldrich, 99%) was suspended in octadecene (5 ml), dried at 100°C for 30 min, and mixed with oleic acid (0.5 ml, vacuum-dried at 100°C) and oleylamine (0.5 ml vacuum-dried at 100°C). When PbBr$_2$ was dissolved, the reaction mixture was heated up to 180°C and preheated caesium oleate in octadecene (0.4 ml, 0.125 M) was injected. The reaction mixture was cooled immediately with an ice bath to room temperature.

**Purification and size-selection of CsPbBr$_3$ nanocrystals**. A critical factor for self-assembly of cubic-shaped CsPbX$_3$ NCs is to start with an initially high level of monodispersity. The crude solution was centrifuged at 12100 rpm for 5 min, following which the supernatant was discarded, and the precipitate was dissolved in 300 µl hexane. The hexane solution was centrifuged again and the precipitate was discarded. The supernatant was diluted two times and used for further purification. Subsequently, two methods of purification of the NCs were applied: (a) 50 µl hexane, 0.6 µl oleic acid, and 0.6 µl oleylamine were added to 50 µl NCs in hexane. The colloid was destabilized by adding 50 µl acetone, followed by centrifuging and dispersing the NCs in 300 µl toluene. This solution was used further for the preparation of the 3D-superlattices. (b) 50 µl hexane and 100 µl toluene were added to 50 µl NCs in hexane. The colloid was destabilized by adding 50 µl acetonitrile, followed by centrifuging and dispersing the NCs in 300 µl toluene. This solution was used further for the preparation of the 3D-superlattices.

**Preparation of 3D-superlattices**.

CsPbBr$_3$ NC superlattices were prepared on glass or on 5 × 7 mm silicon substrates. Shortly before the self-assembly process, the silicon substrate was dipped into 4% solution of HF in water for 1 min, followed by washing with water. In a typical assembly process, the substrate was placed in a 10 × 10 mm Teflon well and 10 µl of purified NCs in toluene were spread onto the substrate. The well was covered with a glass slide and the toluene was then allowed to evaporate slowly. 3D-superlattices of CsPbBr$_3$ NCs were formed upon complete evaporation of the toluene. Typical lateral dimensions of individual superlattices ranged from 1 to 10 µm wherein some of them arrange into clusters of several superlattices and others remain spatially well-isolated so that PL measurements can be performed on an individual superlattice.

More intense purification or greater polydispersity of NCs led to disordered or 2D assemblies (glassy films). Furthermore, the formation of NC superlattices can serve to further narrow the size distribution and shape uniformity within the ensemble (with smaller or larger NCs being repelled from the NC domain), especially in the case of simple cubic packing of cubes, which is particularly intolerant to size and shape variations.

B) CHARACTERIZATION METHODS

For PL, time-resolved PL, and second-order photon-correlation measurements on single superlattices, the sample was mounted on *xyz* positioning stages and excited with a fibre-coupled excitation laser either in continuous wave mode or pulsed mode with 40 MHz repetition rate (pulse duration 50 ps). The excitation laser was filtered with a short-pass filter and directed towards the long-working distance 100× microscope objective (numerical aperture $NA = 0.7$) by a dichroic beam splitter, resulting in a nearly Gaussian-shaped excitation spot with $1/e^2$ radius of 1.4 µm. The emission was collected via the same microscope objective and filtered using a tuneable bandpass filter. For PL measurements, the collected light was then dispersed by a 300 lines/mm grating inside a 750 mm monochromator and detected by an EMCCD camera. For measurements of the PL decay, we filtered the emission with a tuneable band-pass (FWHM = 15 nm) and recorded the decay with an avalanche photo diode single photon detector with a time resolution of $30\,\text{ps}$ connected to a time-correlated single-photon-counting system. The photon correlation was recorded using a similar setup with two detectors in a Hanbury–Brown–Twiss setup configuration.

To record streak camera images and first-order coherence measurements, we excited the sample with a frequency-doubled regenerative amplifier seeded with a mode-locked Ti:sapphire laser with a pulse duration of 100–200 fs and a repetition rate of 1 kHz. For both excitation and detection, we used an 80 mm lens ($NA = 0.013$ after iris), resulting in an excitation spot area of 20 × 40 µm. The recorded PL was dispersed by a grating with 150 lines/mm in a 300 mm spectrograph and detected with a streak camera with a nominal time resolution of 2 ps. First-order coherence measurements were performed using a Michelson interferometer where a non-polarizing beam splitter is used to split and recombine the light in the two interferometer arms, with one arm including a retroreflector on a delay stage with 100 nm step resolution. A tuneable band-pass filter (FWHM = 15 nm) is applied to select the emission from either the coupled or the uncoupled QDs. The interferogram was recorded as real-space images of the recombined and focused detection beams on a camera.

## II) OPTICAL PROPERTIES OF SUPERFLUORESCENCE

### A) SUPERRADIANCE, SUPERFLUORESCENCE, AND SUBRADIANCE

As shown in Figure 3b, we observed that the PL decay of the SF state is initially very fast and cannot be described with a single exponential because the decay rate is dependent on the number of excited TLS, $\Gamma(N) \sim N$, and therefore decreases during the decay. Consequently, the SF decay rate should converge towards the decay rate of the uncoupled nanocrystals. However, we observe that the SF decay trace crosses the bi-exponential PL decay of the uncoupled QDs after 97% of the photons are emitted due to long decay components. These long decay components might originate from coupled QDs where the individual dipoles are out of phase and interfere destructively, known as subradiance (SBR) [28,34]. In ensembles with inhomogeneously broadened PL, SF and subradiant states can coexist, and we find a good agreement of the predicted excited state population with the measured PL decay[35].

An out-of-phase coupling amongst the QDs is expected to result in a higher photon energy of the subradiant state compared to the SF state. In Extended Data Figure 2, we provide an analysis of the dynamical energy shift observed at high excitation power density. Examples of emission spectra as a function of time are reported in Extended Data Figure 2a. In Extended Data Figure 2b, we plot the fitted centre photon energy of time-sliced PL spectra (2 ps bin) as a function of the fitted peak area (*i.e.*, the time-dependent emission intensity), as obtained from excitation power-dependent streak camera images. This effectively shows the energetic shift of the SF state as a function of its occupation, with the different curves representing different initial excitation powers. The green arrows indicate the time sequence of the individual analysed spectral traces. By increasing the excitation power, we observe that the initial dynamical red-shift is the largest for the highest excitation power, as is expected from its relationship to the number of excited coupled QDs. Hence, when the number of excited coupled QDs decreases during the decay process, the emission energy blue-shifts to higher energy. We observe the most pronounced energetic blue-shift for the highest excitation power, resulting in a final emission with a photon energy that has been boosted incrementally more in comparison to the blue-shift for low excitation power, which is another indication of the presence of subradiant states that emit at higher energies. For high excitation power, the SF state becomes depopulated much faster since more QDs are coupled simultaneously.

Then, at long timescales after the initial decay, the percentage of subradiant states becomes dominant, resulting in a blue-shift of the PL emission.

B) SUPERFLUORESCENCE FIT MODEL

SF decay traces as in Figure 3b cannot be fitted well with mono- or bi-exponential functions because the decay rate is proportional to the number of excited coupled QDs $\Gamma(N) \sim N$, which also decays over time. Furthermore, the resulting characteristic decay neither exactly follows stretched-exponential nor a power-law dependence[36], whereas the PL decay of the uncoupled QDs is well described by a bi-exponential behaviour, where the initial fast decay $\tau_{QD} = 349.8 \pm 0.4$ ps accounts for over 96% the total emitted photons. Nevertheless, we found that the best approximate fit to the SF decay trace is the Kohlrausch stretched-exponential decay model $\frac{f(t)}{f(0)} = \exp[-(\Gamma_{\text{stretched}} \cdot t)^\beta]$, where $\Gamma_{\text{stretched}}$ is the average decay time and $\beta \in [0,1]$ is the stretch parameter, which represents the distribution of decay rates[37]. Using this model to fit the SF decay curve, we obtain an average decay time $\tau_{\text{stretched}} = 40.4 \pm 0.5$ ps and a stretch parameter $\beta = 0.457 \pm 0.002$.

At a high excitation density, as shown in Figure 5b, we observe oscillations in the decay. To model the SF decay with this characteristic ringing behaviour, we used a decay model consisting of a bi-exponential decay that is multiplied by a damped oscillating term $1 + B \cdot \exp(-\gamma_{\text{Damp}} t) \cdot \cos(\omega t + \phi_0)$. Furthermore, for the rising edge of the emitted pulse we take into account a Gaussian rise term $\sim \exp[-\left(\frac{t-\tau_D}{\tau_{\text{rise}}}\right)^2]$, such that the complete fit function is given by (ref. [38]):

$$f(t) = \sum_{n=1,2} A_n \cdot \exp\left(\frac{\tau_{\text{rise}}^2}{4\tau_n^2} - \frac{t-\tau_D}{\tau_n}\right)$$

$$\cdot \left[\frac{1}{2}\left(1 + B \cdot \exp(-\gamma_{\text{Damp}}(t - \tau_D)) \cdot \cos(\omega(t - \tau_D) + \phi_0)\right)\right.$$

$$\left.\cdot \left[1 + \text{erf}\left(\frac{t-\tau_D}{\tau_{\text{rise}}} - \frac{\tau_{\text{rise}}}{2\tau_n}\right)\right]\right]$$

Here, $A_n$ are the amplitudes of the exponential decay with the corresponding decay time constants, $\tau_n$. Both the fast decay time, shown in Figure 5c, and the long decay time component (Extended Data Figure 2d and e) decrease upon increasing the excitation density, whereas the rise time, $\tau_{\text{rise}} = 3.44 \pm 1.05$ ps, stays approximately constant (probably clamped by the time resolution of the setup). In the upper panel of Figure 5c, we plot the power-dependent effective decay time $\tau_{\text{SF}} = \frac{A_1\tau_1 + A_2\tau_2}{A_1 + A_2}$, where $\tau_1$, $\tau_2$ are the decay times of the bi-exponential fit and $A_1$, $A_2$ the corresponding amplitudes, which was fitted with $\tau_{\text{SF}}(P) = \frac{\tau_{\text{QD}}}{\zeta \cdot P + 1} + y_0$. We obtain good agreement with the expected behaviour ($\tau_{\text{SF}} \sim \tau_{\text{QD}}/N$) for a value of $\zeta = 0.148 \pm 0.004 \frac{cm^2}{\mu J}$. In the lower panel of Figure 5c, we plot the delay time $\tau_D$ as a function of the excitation power. In our analysis, the delay time is composed of the actual delay time due to the SF build-up and a systematic, constant time-offset because the absolute arrival time of the excitation pulse (which has a different wavelength than the emission) at the sample cannot be measured reliably at the required precision from the streak camera data. We observe a decrease in $\tau_D$ of ~4 ps when increasing the excitation density by almost 2 orders of magnitude. We have fitted this behaviour with $\tau_D = y_{\text{offset}} + A \cdot \ln(\zeta P_{\text{Exc}} + 1)/(\zeta P_{\text{Exc}} + 1)$ because we assume that $\tau_D \sim \ln(N)/N$ and that the number of excited coupled emitters $N \sim \alpha P_{\text{Exc}} + 1$ is proportional to the excitation power. Herein, we use a fixed value $\zeta = 0.148 \pm 0.004 \frac{cm^2}{\mu J}$, which we obtained from the fit of the effective decay in the upper panel of Figure 5c. The resulting fit agrees very well with the data. To obtain the absolute time delay, we subtracted the constant offset of the time-delay fit from the time-delay data points. SF occurs when $\sqrt{\tau_{\text{SF}}\tau_D} < T_2^*$, where $T_2^*$ is the exciton pure dephasing time. Considering that the coherence time $T_2 < T_2^*$ extracted from the full-width at half-maximum of single QDs[10] is of the order of $T_2 = 6.6$ ps, our measurements reveal a fast decay of ~8 ps and a delay time of < 1 ps which satisfies the criterion for the appearance of SF.

## Additional references

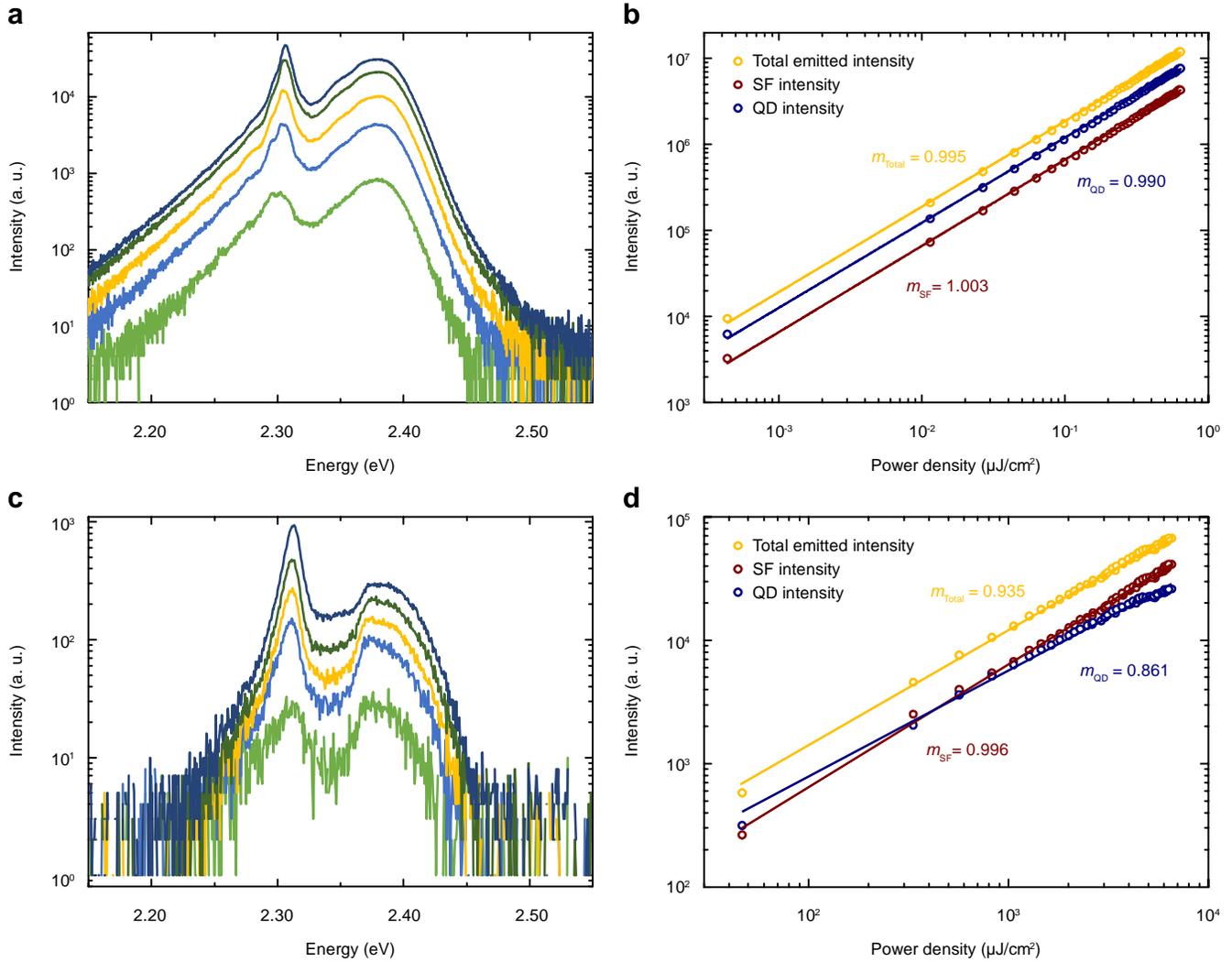

**Extended Data Figure 1 | Power dependent PL properties. a**, Colour-coded PL emission of a single superlattice in the low-power excitation regime, shown for increasing excitation fluence of 10 nJ/cm² (light green), 60 nJ/cm² (light blue), 150 nJ/cm² (yellow), 310 nJ/cm² (dark green) and 600 nJ/cm² (dark blue). **b**, PL intensity integrated over the spectral emission range of the uncoupled QDs (blue circles) and coupled QDs (red circles) in a log-log plot and the total emitted intensity (yellow circles). Fits to the data reveal a perfectly linear behaviour, as represented by a fitted power-law exponent of 1. **c**, Colour-coded PL emission of a single superlattice in the high-power excitation regime, shown for increasing excitation fluence of 330 µJ/cm² (light green), 1270 µJ/cm² (light blue), 2130 µJ/cm² (yellow), 3470 µJ/cm² (dark green) and 6330 µJ/cm² (dark blue). **d**, PL intensity integrated over the spectral emission range of the uncoupled QDs (blue) and coupled QDs (red) in a log-log plot and the total emitted intensity (yellow). Fits to the data reveal a power-law behaviour with a linear increase for the SF emission, a slightly sublinear increase for the uncoupled QDs and a less sublinear increase for the total emitted intensity.

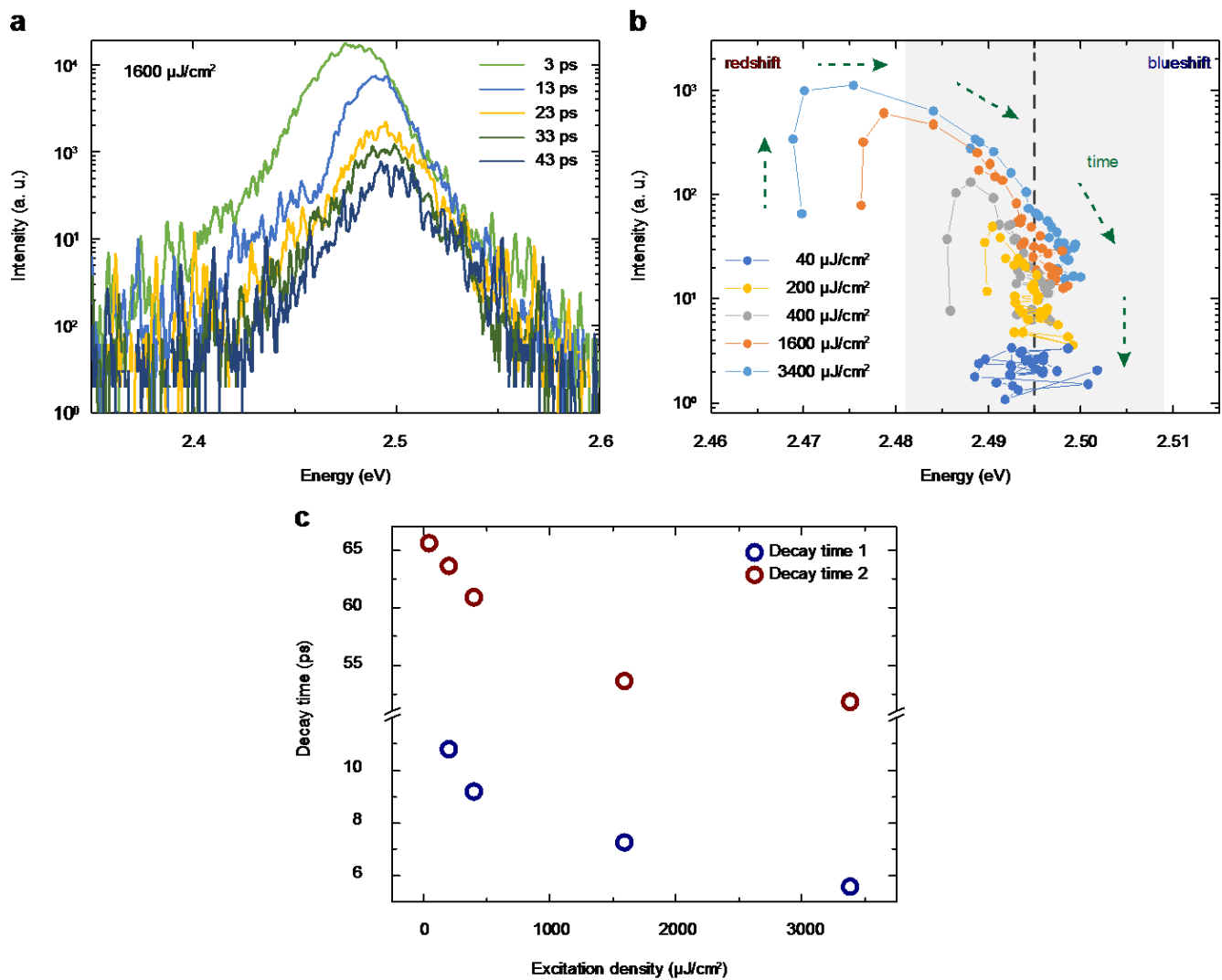

**Extended Data Figure 2 | SF decay and energetic redshift. a**, PL spectra (integrated over 2 ps time window) at different time delays in a semi-log scale. **b**, Time-integrated PL spectra are fitted to a single Gaussian peak function. The fitted peak amplitude as a function of the emission energy is plotted for various excitation densities. Green arrows indicate the time evolution of the emission peak. **c**, Fast and slow PL decay time components $\tau_1$ and $\tau_2$ of the SF bi-exponential fit model as a function of excitation density.

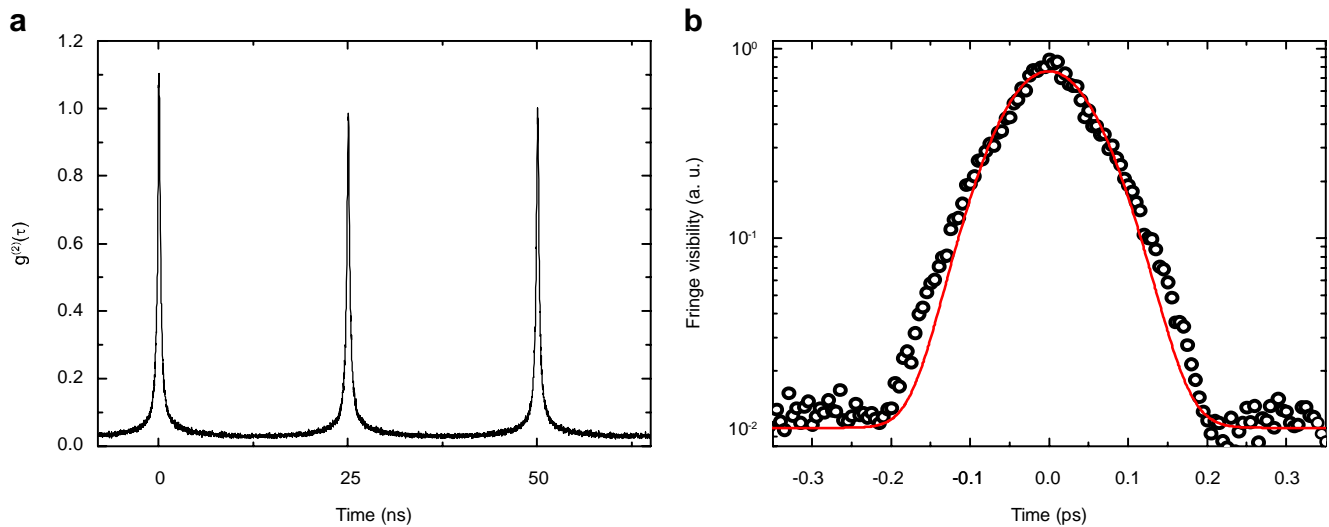

**Extended Data Figure 3 | Photon bunching in pulsed excitation and Gaussian first-order coherence decay. a**, Second-order photon correlation measurement of the SF peak showing photon bunching at zero delay under pulsed excitation with a 40 MHz repetition rate. **b**, First-order coherence extracted from the fringe contrast of the interferograms as a function of differential delay time between the arms of a Michelson interferometer, revealing a mixture of Gaussian (Kubo) and exponential decay (for some of the superlattices).

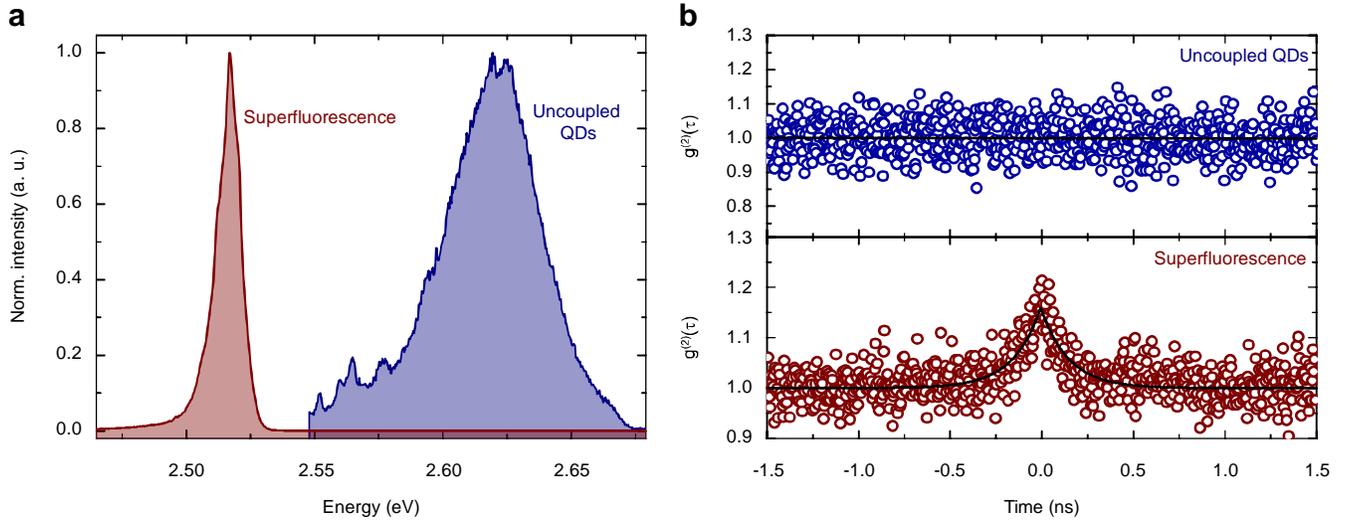

**Extended Data Figure 4 | Superfluorescence in CsPbBr$_2$Cl quantum dot superlattices. a**, Band-pass filtered PL spectra of uncoupled QDs (blue) and SF emission (red) of CsPbBr$_2$Cl perovskite superlattices. **b**, Second-order photon correlation measurement of uncoupled QDs (top panel, $g^{(2)}(\tau) = 1$) showing flat correlation function, and the SF emission peak band (lower panel, $g^{(2)}(0) = 1.15$) showing photon bunching.